\pdfoutput=1

\newcommand{\ALRO}{\ch{Ag3LiRh2O6}}
\newcommand{\LRO}{\ch{Li2RhO3}}

\newcommand{\Tn}{$T_{\text {N}}$}
\newcommand{\Tw}{$\Theta_{\text {CW}}$}
\newcommand{\uf}{$\mu_{\text {eff}}$}
\newcommand{\ub}{$\mu_{\text {B}}$}

\documentclass{nature}

\usepackage[version=3]{mhchem} 
\usepackage[T1]{fontenc}       
\usepackage[version=3]{mhchem} 
\usepackage[T1]{fontenc}       
\usepackage{graphicx}          
\usepackage{dcolumn}           
\usepackage{bm}                
\usepackage{xcolor}            
\usepackage{colortbl}
\usepackage{amsmath}           
\usepackage{hyperref}          
\usepackage{tabularx}
\usepackage{lineno}
\usepackage{chemformula} 
\usepackage{xcolor,colortbl}
\usepackage{multirow}
\usepackage{lineno}
\title{
\begin{center}
{Pressure tuning of competing interactions on a honeycomb lattice}
\end{center}
}
\author{
Piyush~Sakrikar$^{1}$\footnote{These authors contributed equally to this work.},
Bin~Shen$^{2*}$,
Eduardo~H.~T.~Poldi$^{3,4*}$,
Faranak~Bahrami$^{1}$,
Xiaodong~Hu$^1$,
Eric~M.~Kenney$^{1,5}$,
Qiaochu Wang$^6$,
Kyle~W.~Fruhling$^1$,
Chennan~Wang$^7$, 
Ritu~Gupta$^7$,
Rustem~Khasanov$^7$,
Hubertus~Luetkens$^7$,
Stuart~A.~Calder$^8$,
Adam~A.~Aczel$^8$,
Gilberto~Fabbris$^3$,
Russell~J.~Hemley$^4$,
Kemp~W.~Plumb$^6$,
Ying~Ran$^1$,
Philipp~Gegenwart$^2$,
Alexander~A.~Tsirlin$^2$,
Daniel~Haskel$^3$,
Michael~J.~Graf$^1$,
Fazel~Tafti$^1$
}
\begin{document}
\maketitle

\begin{affiliations}
 \item{Department of Physics, Boston College, Chestnut Hill, MA 02467, USA}
 \item{Experimental Physics VI, Center for Electronic Correlations and Magnetism, University of Augsburg, 86159 Augsburg, Germany}
 \item{Advanced Photon Source, Argonne National Laboratory, Argonne, IL 60439, USA}
 \item{Departments of Physics and Chemistry, University of Illinois Chicago, Chicago, IL 60607, USA}
 \item{Muon Science Laboratory, Institute of Materials Structure Science, High Energy Accelerator Research Organization (KEK), Tsukuba, Ibaraki 305-0801, Japan}
  \item{Department of Physics, Brown University, Providence, RI 02912, United States}
 \item{Laboratory for Muon Spin Spectroscopy, Paul Scherrer Institute, CH-5232 Villigen, Switzerland}
 \item{Neutron Scattering Division, Oak Ridge National Laboratory, Oak Ridge, TN 37831, USA}
\end{affiliations}

\begin{abstract}
Magnetic exchange interactions are mediated via orbital overlaps across chemical bonds.
Thus, modifying the bond angles by physical pressure or strain can tune the relative strength of competing interactions.
Here we present a remarkable case of such tuning between the Heisenberg ($J$) and Kitaev ($K$) exchange, which respectively establish magnetically ordered and spin liquid phases on a honeycomb lattice.
We observe a rapid suppression of the N\'{e}el temperature (\Tn) with pressure in \ALRO, a spin-1/2 honeycomb lattice with both $J$ and $K$ couplings.
Using a combined analysis of x-ray data and first-principles calculations, we find that pressure modifies the bond angles in a way that increases the $|K/J|$ ratio and thereby suppresses \Tn.
Consistent with this picture, we observe a spontaneous onset of muon spin relaxation ($\mu$SR) oscillations below \Tn\ at low pressure, whereas in the high pressure phase, oscillations appear only when $T< T_{\textrm{N}}/2$.
Unlike other cadidate Kitaev materials, \ALRO\ is tuned toward a quantum critical point by pressure while avoiding a structural dimerizaion in the relevant pressure range.
\end{abstract}


\section*{\label{sec:intro}Introduction}
Materials with a honeycomb lattice and heavy elements can sustain anisotropic Kitaev interactions which favor a quantum spin liquid (QSL) ground state~\cite{jackeli_mott_2009,chaloupka_kitaev-heisenberg_2010,takagi_concept_2019}.
The same materials also host isotropic Heisenberg interactions which favor a long-range magnetic order (LRO)~\cite{singh_relevance_2012,chaloupka_zigzag_2013}.
Theoretically, the QSL ground state could be established by tuning the competition between the Kitaev and Heisenberg interactions in favor of the former~\cite{rau_generic_2014,kimchi_kitaev-heisenberg_2014}.
One approach to this problem would be to chemically design new materials with a large Kitaev to Heisenberg coupling ratio $|K/J|$.
Unfortunately, this is proven to be an extremely challenging task~\cite{winter_challenges_2016,winter_models_2017,bahrami_effect_2021,kenney_coexistence_2019,kitagawa_spinorbital-entangled_2018}.
An alternative approach would be to use external parameters such as magnetic field strength~\cite{banerjee_excitations_2018} or angle~\cite{tanaka_thermodynamic_2022} to tune an existing material away from the Heisenberg limit and toward the Kitaev limit.
In this work, we present a successful case of such tuning by applying hydrostatic pressure, instead of magnetic field, on the honeycomb lattice of \ALRO.

\ALRO\ is synthesized from the parent compound \LRO\ by replacing the small inter-layer Li atoms with large Ag atoms in a topotactic exchange reaction (Fig.~\ref{fig:MAG}a)~\cite{bahrami_first_2022}.
Changing the interlayer atoms induces a trigonal distortion in \ch{RhO6} octahedra, which enhances the Ising-like anisotropy of the pseudospin-1/2 states~\cite{bahrami_first_2022}. 
As a result, a robust antiferromagnetic (AFM) order is established in \ALRO\ at \Tn$=$100~K, in stark contrast to the glassy transition at 6~K in \LRO.
The large \Tn\ in \ALRO\ indicates a dominant Heisenberg interaction, i.e. a small $|K/J|$ ratio.
We decided to study this material under pressure based on quantum chemistry calculations that predict the $|K/J|$ ratio could be increased by modifying the $\angle$Rh-O-Rh bond angles within the honeycomb layers (Fig.~\ref{fig:MAG}b)~\cite{katukuri_strong_2015}.
$4d$ transition metal systems such as \LRO\ and \ALRO\ are particularly sensitive to changes of bond angles, since they have comparable spin-orbit coupling and crystal field energy scales~\cite{katukuri_electronic_2014}.

Our multiprobe investigations reveal three pieces of evidence for a shift in the balance between the Heisenberg and Kitaev interactions with increasing pressure in \ALRO.
(i) Magnetization measurements show a rapid suppression of \Tn\ under pressure up to 3~GPa, beyond which, the AFM order disappears. 
(ii) X-ray diffraction (XRD) confirms the absence of structural transitions up to 6~GPa, beyond which, the honeycomb lattice undergoes a dimerization transition.
(iii) $\mu$SR experiments reveal a long-range order below \Tn\ at low pressures but a short-range order at high pressures, which becomes long-range only when $T$$<$\Tn$/2$.
Thus, the $\mu$SR data indicates an extended temperature regime of fluctuating short-range magnetism.
Details of the magnetization, XRD, and $\mu$SR data are presented below.

\section*{\label{sec:magnetism}Magnetization}
\begin{figure*}[t]
 \centering
 \includegraphics[width=\textwidth]{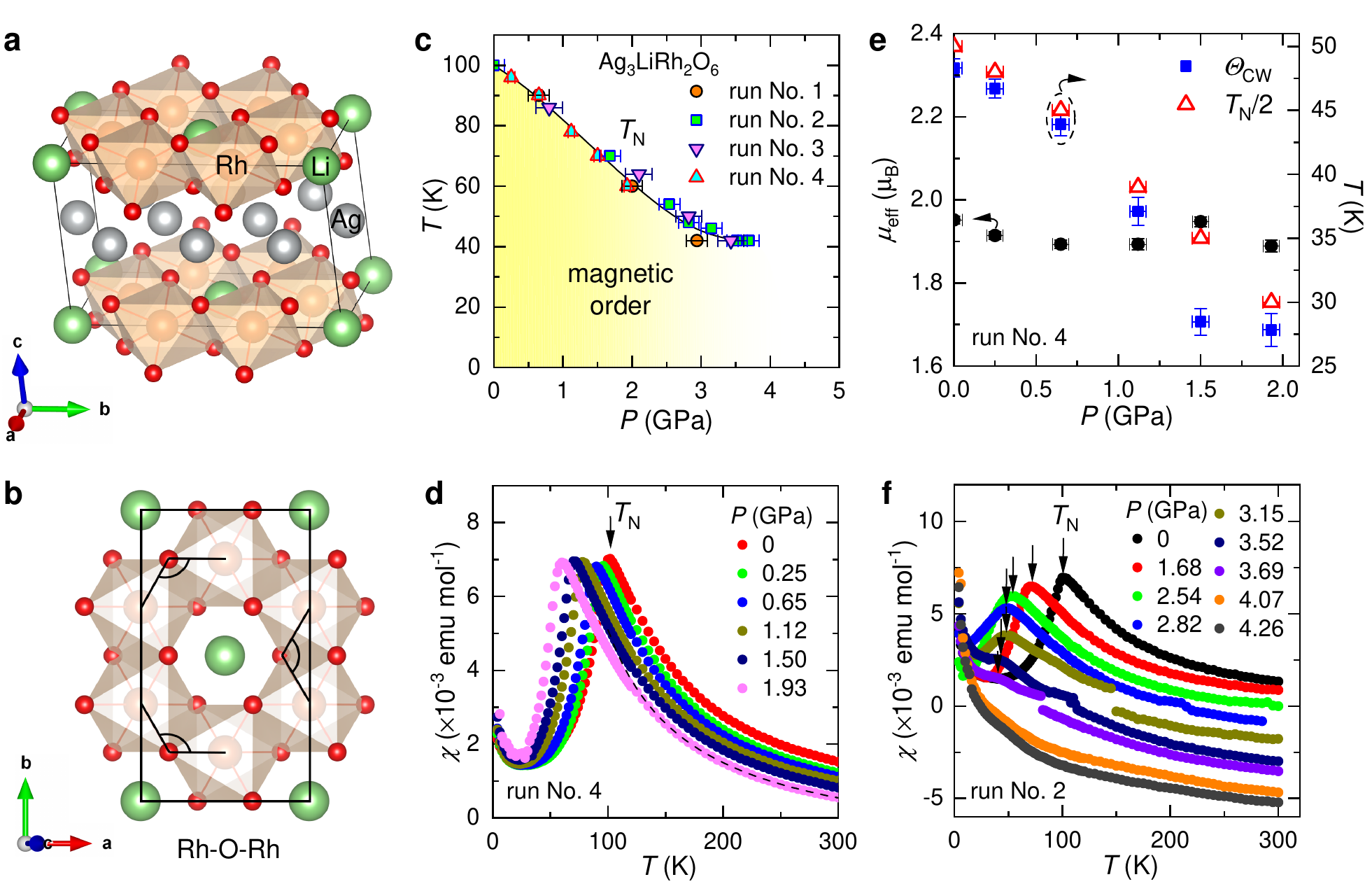}
 \caption{\label{fig:MAG}
 \fontfamily{phv}\selectfont\small{
 \textbf{Magnetization data.}
 (\textbf{a}) Unit cell of \ALRO\ in the monoclinic space group $C2/m$ with Ag atoms between the [LiRh$_2$O$_6$] honeycomb layers.
 (\textbf{b}) $\angle$Rh-O-Rh bond angles within a honeycomb layer.
 (\textbf{c}) Suppression of \Tn\ with increasing pressure.
 (\textbf{d}) \Tn\ is identified by the peak in $\chi(T)$ at different pressures.
 (\textbf{e}) Both \Tn\ and \Tw\ decrease in parallel with pressure while \uf\ remains nearly unchanged. 
 All data in this panel are from run No. 4.
 (\textbf{f}) The peak in susceptibility data (\Tn) disappears at $P>4$~GPa.
 }}
\end{figure*}
We started the high pressure investigations of \ALRO\ by measuring DC magnetic susceptibility of a polycrystalline sample inside a ceramic anvil pressure cell.
To reach the maximum pressure of about 5.5~GPa, we used a pair of anvils with small culets in runs 1, 2 and 3 (Methods).
To obtain higher quality data for the Curie-Weiss (CW) analysis, we used another pair of anvils with larger culets which limited the pressure to 2~GPa in run 4. 

The first observation in Fig.~\ref{fig:MAG}c is a rapid suppression of \Tn\ with pressure at a rate of $-20$~K/GPa up to about 3~GPa.
At each pressure, \Tn\ was obtained from the peak in the $\chi(T)$ curve as seen in Fig.~\ref{fig:MAG}d for run 4.
The high quality of these data enabled us to perform Curie-Weiss (CW) fits to extract the CW temperature (\Tw) and effective magnetic moment (\uf).  
Plotting \Tn, \Tw, and \uf\ as a function of pressure in Fig.~\ref{fig:MAG}e reveals a parallel suppression of \Tn\ and \Tw\ with pressure, while \uf\ remains nearly unchanged.
Since \Tw\ is proportional to the Heisenberg coupling $J$, the parallel suppression of \Tw\ and \Tn\ indicates a weakening of the average $J$ under pressure.
The value of \uf~$\approx 1.9$~\ub, which is unaffected by pressures, is close to the expected moment for a pseudospin-1/2 state.
In the supplementary information, we also provide DFT results that confirm the robustness of the pseudospin-1/2 state up to 5~GPa.
These observations suggest that while the pseudospin-1/2 state in \ALRO\ remains unchanged under pressure, the Heisenberg interactions weaken with increasing pressure, resulting in a rapid suppression of \Tn.

Switching to anvils with smaller culet sizes, we extended measurements of $\chi(T)$ to higher pressures in runs 1, 2, and 3 (Fig.~\ref{fig:MAG}f and the supplementary Fig.~S1). 
The $\chi(T)$ curves qualitatively changed at $P>3$~GPa, where the AFM peak became smaller in magnitude and nearly disappeared at $P>4$~GPa (Fig.~\ref{fig:MAG}f).
The disappearance of the AFM peak at high pressures suggests that the Kitaev coupling $K$ is suppressed at a slower rate than Heisenberg coupling $J$, hence the ratio $|K/J|$ is increased with increasing pressure.

Other than the $J$-$K$ model discussed above, an alternative theoretical framework for a honeycomb lattice with spin-1/2 particles would be the $J_1$-$J_2$ model~\cite{mulder_spiral_2010,gong_phase_2013,bishop_complete_2012,ganesh_deconfined_2013}.
Such a model is particularly relevant for \ALRO\ due to the Ising anisotropy of its pseudospin-1/2 state.
The $J_1$-$J_2$ model is frustrated when $J_2$ is AFM ($J_2<0$), regardless of the sign of $J_1$.
In such a model, $\Theta_{\textrm{CW}}\propto \frac{J_1+2J_2}{3}$ is positive and decreases with pressure if $J_1\ll -J_2>0$ (i.e. with FM $J_1$ and AFM $J_2$).
Regardless of using the $J$-$K$ model or $J_1$-$J_2$ model, the magnetization data presented in Fig.~\ref{fig:MAG} are consistent with competing interactions.

\section*{\label{sec:xrd}X-ray diffraction}
We performed XRD measurements under pressure with two goals in mind.
First, to confirm that the suppression of the AFM order was not due to a structural transition, and second, to correlate the \Tn\ suppression with a change of $\angle$Rh-O-Rh bond angle.

Our search for a pressure-induced structural transition was motivated by previous studies on the hyper-honeycomb system $\beta$-\ch{Li2IrO3}, which similar to \ALRO, has a high \Tn\ of 38~K at ambient pressure and loses its AFM order under pressure~\cite{takayama_hyperhoneycomb_2015,veiga_pressure_2017,majumder_breakdown_2018,shen_interplay_2021}.
However, unlike in \ALRO, \Tn\ remains nearly independent of pressure in $\beta$-\ch{Li2IrO3} until the AFM order disappears abruptly at $P_c=1.4$~GPa~\cite{shen_interplay_2021,majumder_breakdown_2018}.
The sudden loss of the AFM order in $\beta$-\ch{Li2IrO3} is unrelated to competing interactions.
Instead, it originates form the loss of local moments due to the formation of Ir$_2$ dimers under pressure~\cite{veiga_pressure_2017,shen_interplay_2021,majumder_breakdown_2018}.
Measurements of x-ray magnetic circular dichroism (XMCD)~\cite{veiga_pressure_2017} reveal a quenching of both spin and orbital moments due to this dimerization at $P_c=1.4$~GPa. 
Thus, we performed high-pressure x-ray diffraction on \ALRO\ to distinguish between two mechanisms for the loss of AFM order: (i)  structural dimerization, and (ii) competing interactions. 

\begin{figure*}[h]
 \centering
 \includegraphics[width=\textwidth]{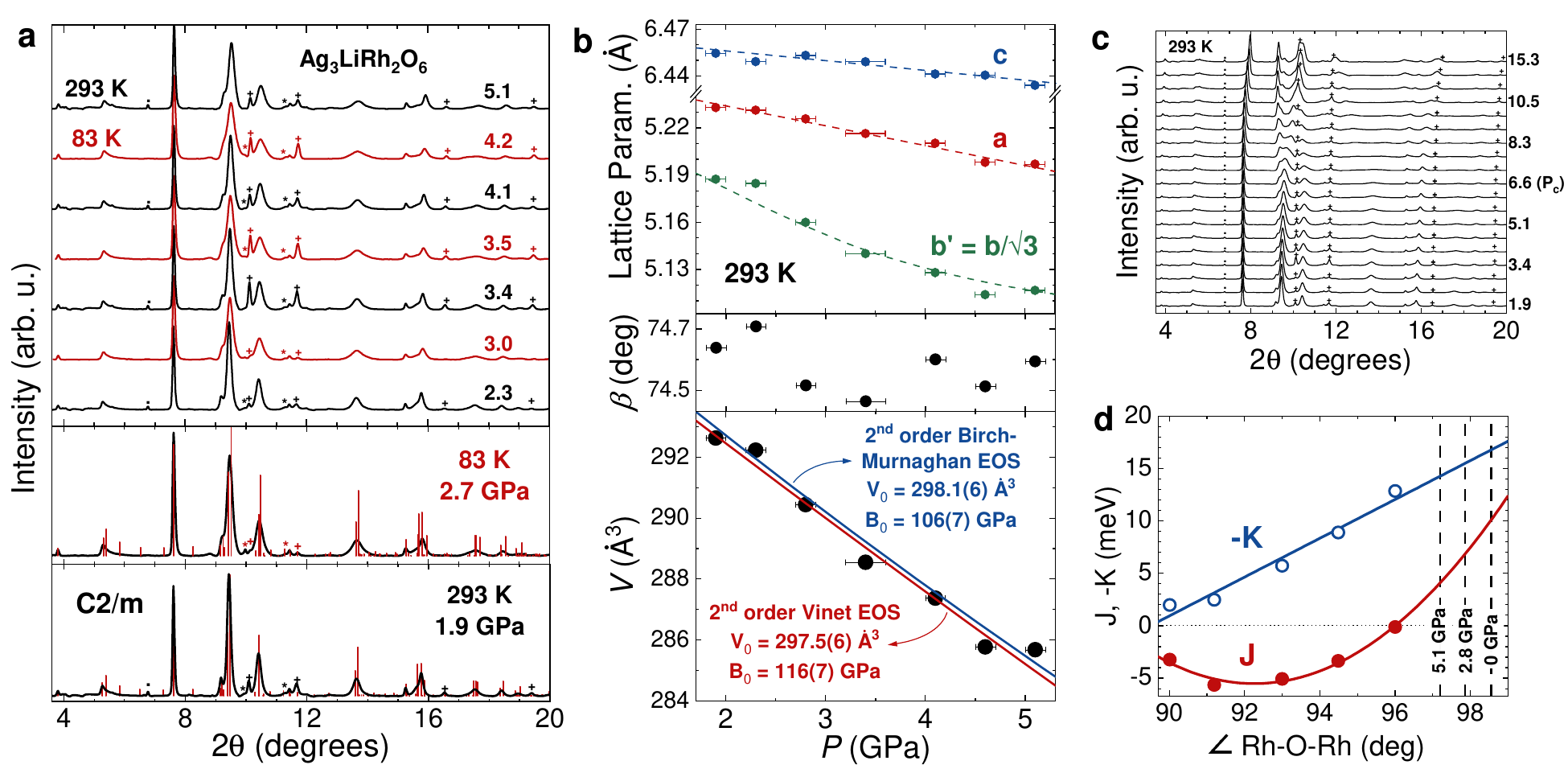}
 \caption{\label{fig:XRD}
 \fontfamily{phv}\selectfont\small{
 \textbf{X-ray diffraction.}
 (\textbf{a}) Pressure-dependent diffraction patterns at 293~K (black) and 83~K (red). 
 The $*$, + and . symbols indicate, respectively, the Re peaks (gasket), Au peaks (manometers), and boron carbide seat. 
 Red vertical bars in the bottom panel indicate calculated Bragg peak intensities at 1.9~GPa (293~K) and 2.7~GPa (83~K). 
 (\textbf{b}) Monoclinic unit cell parameters, angle $\beta$, and volume plotted as a function of pressure. 
 The $P$-$V$ data are fitted (solid lines) using both 2nd order Vinet and 2nd order Birch-Murnaghan equations of state, rendering comparable values for the bulk modulus ($B_0$) and ambient pressure volume ($V_0$).
 (\textbf{c}) The bifurcation of the $9.5^\circ$ Bragg peak at $P_c=6.6$~GPa indicates a dimerization transition (see also Fig.~S2).
 (\textbf{d}) The linear and quadratic dependence of $K$ and $J$ on $\angle$Rh-O-Rh are reproduced from Ref.~\cite{katukuri_strong_2015}.
 Dashed lines indicate the average bond angle at different pressures. 
 }}
\end{figure*}

The XRD patterns in Fig.~\ref{fig:XRD}a show that the monoclinic $C2/m$ structure is preserved in \ALRO\ from 0 to 5~GPa at both 293~K and 85~K.
Using LeBail fits to these data, we trace the evolution of the unit cell parameters with pressure in Fig.~\ref{fig:XRD}b.
All lattice parameters are smoothly decreasing with increasing pressure, and the monoclinic angle $\beta$ fluctuates around 74.6(1) degrees. 
The absence of a structural transition up to 5~GPa in Figs.~\ref{fig:XRD}a,b rules out the dimerization of Rh$_2$ units as the mechanism of \Tn\ suppression.
This is consistent with the pressure independent \uf\ in Fig.~\ref{fig:MAG}e, since the formation of Rh$_2$ dimers would have quenched the local moments. 

Fig.~\ref{fig:XRD}c shows that a structural transition finally occurs at $P_c=6.6(5)$~GPa, well above the pressure range of \Tn\ suppression in Fig.~\ref{fig:MAG}c.
The structural transition is signaled by a bifurcation of the Bragg peak at 9.5$^\circ$ in Fig.~\ref{fig:XRD}c (see also supplementary Fig.~S2).
A similar dimerization transition has been reported in \ch{Li2RhO3} at nearly the same critical pressure~\cite{hermann_pressure-induced_2019}.

The known presence of stacking faults in \ALRO~\cite{bahrami_first_2022} and limited angular range of the high-pressure XRD data made Rietveld refinements of atomic positions challenging.
Instead, we used the lattice parameters from XRD as input to a density functional theory (DFT) code and found the atomic positions that minimized the free energy (supplementary information).
Using the atomic coordinates from DFT, we evaluated the $\angle$Rh-O-Rh bond angles at high pressures.
The three dashed lines in Fig.~\ref{fig:XRD}d indicate the average values of $\angle$Rh-O-Rh at different pressures, overlaid on a plot of $J$ and $K$ couplings versus $\angle$Rh-O-Rh according to quantum chemistry calculations in \ch{Li2RhO3}~\cite{katukuri_strong_2015}.
The key observation is that $|K/J|$ ratio increases rapidly with increasing pressure as the bond angles approach the critical value of 96$^\circ$ where $J\to0$.
Note that $J$ changes quadratically with bond angle while $|K|$ changes linearly.
This leads to the rapid increase of $|K/J|$ from 1.6 to 2.6 and 3.6 as the pressure increases from 0 to 2.8 and 5.1~GPa, respectively.
Such enhancement of the $|K/J|$ ratio in the absence of a structural transition before 5.5~GPa suggests that competing interactions are responsible for the \Tn\ suppression and disappearance of the AFM peak in Fig.~\ref{fig:MAG}.

We used the calculated $J$ and $K$ curves for \ch{Li2RhO3} in Fig.~\ref{fig:XRD}d, because such calculations do not exist for \ALRO\ at present.
Thus, future material-specific calculations will be necessary for a quantitative analysis.
Nevertheless, the analysis in Fig.~\ref{fig:XRD}d demonstrates how competing Kitaev and Heisenberg interactions could lead to the suppression of the AFM order.
We point out that a similar behavior is expected for competing $J_1$-$J_2$ interactions~\cite{mulder_spiral_2010,gong_phase_2013,bishop_complete_2012,ganesh_deconfined_2013}, although material-specific results have not been reported in this model, unlike the $J$-$K$ model~\cite{katukuri_strong_2015}.

\section*{\label{sec:musr}Muon spin relaxation}
\begin{figure*}[h]
 \centering
 \includegraphics[width=\textwidth]{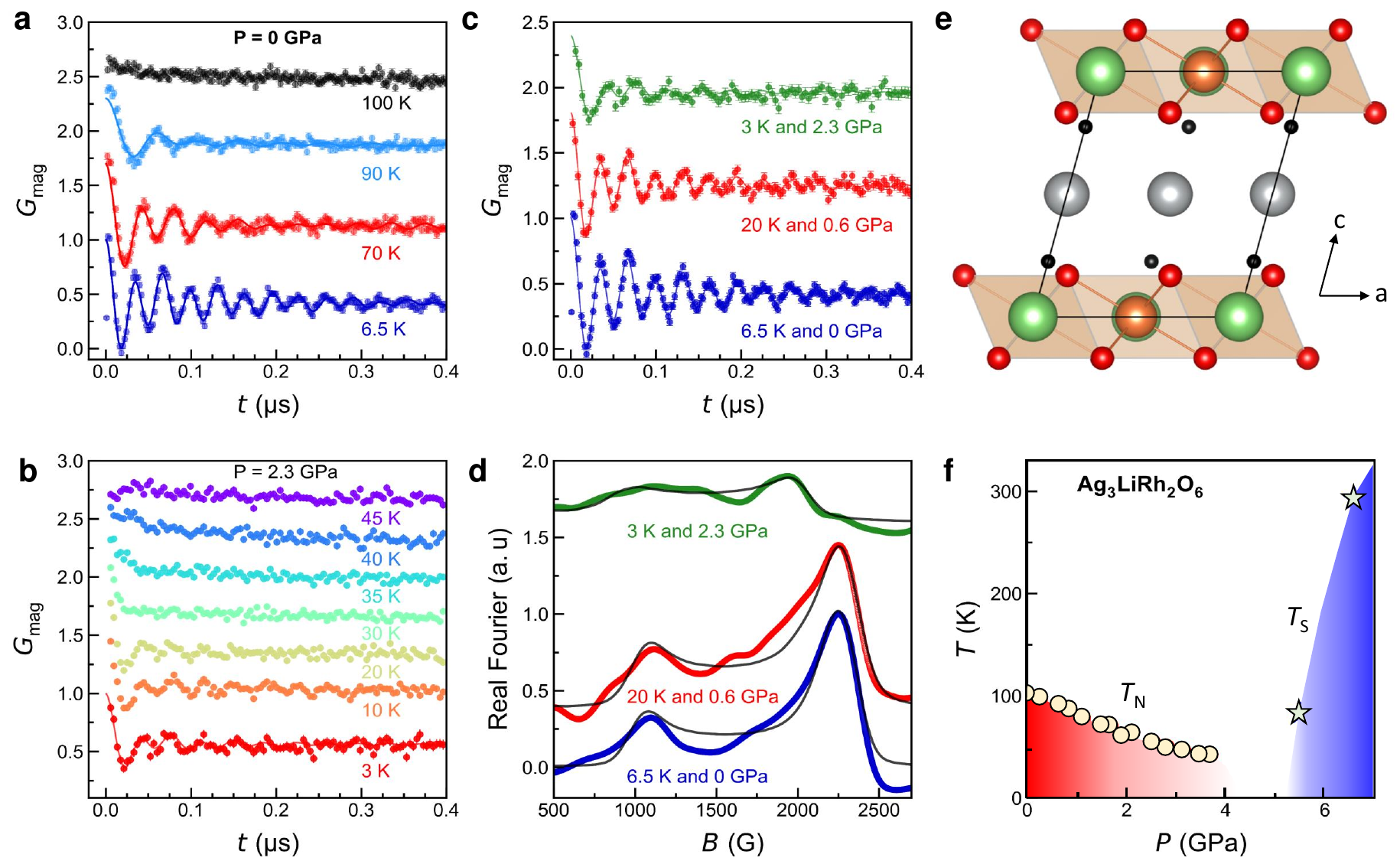}
 \caption{\label{fig:MUSR}
 \fontfamily{phv}\selectfont\small{
 \textbf{$\mu$SR data.}
 (\textbf{a}) Muon polarization at a low pressure showing oscillations immediately below \Tn. 
 (\textbf{b}) At a high pressure, oscillations do not appear until \Tn/2. 
 (\textbf{c}) Comparing the low-$T$ polarization curves at low-pressure (0 and 0.6~GPa) and high-pressure (2.3~GPa).
 (\textbf{d}) Comparing the Fourier transforms of polarization curves.
 (\textbf{e}) Visualizing the muon stopping sites (black circles) in the lattice structure.
 (\textbf{f}) Phase diagram of the magnetic (\Tn) and structural ($T_{\mathrm S}$) transitions in \ALRO.
 }}
\end{figure*}

In $\mu$SR, positively charged spin polarized muons are implanted in a sample to probe the local magnetic field at some preferred crystallographic stopping site(s). 
The average time evolution of the muon polarization $G(t)$ is monitored by detection of positrons which are preferentially emitted along the muon polarization direction upon its decay (lifetime $\tau=2.2~\mu$s). 
Long-range magnetic order is signaled by the onset of oscillations in $G(t)$ in zero magnetic field, and decay of $G(t)$ (depolarization) can be caused by either magnetic disorder or dynamical fluctuations.
The polarization curves in Fig.~\ref{fig:MUSR} are labeled $G_{\textrm{mag}}(t)$ to indicate the removal of background signal from the pressure cell~\cite{khasanov_perspective_2022} and a small non-magnetic signal from silver inclusions in the sample from the total polarization signal $G(t)$.
Details of background subtraction are given in the supplementary information (Fig.~S3).

Before presenting any quantitative analysis, we reveal a qualitative difference between $G_{{\textrm{mag}}}(t)$ curves obtained at low-pressure ($P<2$~GPa) and high-pressure ($P>2$~GPa) in Figs.~\ref{fig:MUSR}a,b.
Whereas the oscillations appear immediately below \Tn$=$95~K at $P=0$ (Fig.~\ref{fig:MUSR}a), they do not appear until the temperature is decreased to half the \Tn$=$43~K at $P=2.3$~GPa (Fig.~\ref{fig:MUSR}b).
The observation of spontaneous oscillations below \Tn\ at low pressures (Fig.~\ref{fig:MUSR}a) indicates the onset of long-range ordering.
This is a typical behavior in a sample without magnetic disorder.
The surprising result is that at high pressures (Fig.~\ref{fig:MUSR}b), oscillations associated with a long-range order do not appear until temperatures below 20~K, which is half the \Tn$=$42~K at 2.3~GPa (Fig.~\ref{fig:MAG}c). 
In the intermediate range $\frac{T_{\textrm{N}}}{2}<T<T_{\textrm{N}}$, oscillations are replaced with a fast depolarization, suggesting short-range magnetic ordering.
A similar behavior has been reported at ambient pressure in \ch{Li2RhO3} and $\alpha$-\ch{Li2IrO3} which are proximate Kitaev spin liquid materials~\cite{khuntia_local_2017,choi_spin_2019}. 
Specifically, \LRO\ is proposed to be a proximate Kitaev spin liquid system where disorder establishes a spin glass phase~\cite{katukuri_strong_2015}.
Thus, pressure seems to tune the static magnetism of \ALRO\ toward the dyanmic behavior observed in its parent compound \LRO. 

In addition to the qualitative differences between low-pressure and high-pressure polarization curves at $\frac{T_{\textrm{N}}}{2}< T <T_{\textrm{N}}$, we also find quantitative differences at $T\ll T_{\textrm{N}}$.
Figures~\ref{fig:MUSR}c,d show $G_{\textrm{mag}}(t)$ spectra and their Fourier transforms at ambient pressure, 0.6~GPa, and 2.3~GPa for $T\le 10$~K. 
We fit the magnetic polarization curves to the following expression
\begin{align}
\label{eq:Gmag}
G_{\textrm{mag}}(t)&=f_{\text{osc}}\left[ f_1J_0(\gamma_\mu \Delta B_1t)\cos(\gamma_\mu B_{1,\textrm{avg}}t)\exp(-\Lambda_1t) + f_2J_0(\gamma_\mu \Delta B_2t)\cos(\gamma_\mu B_{2,\textrm{avg}}t)\exp(-\Lambda_2t) \right] \nonumber \\
&+ \left(1-f_{\text{osc}}\right)\exp(-\lambda_Lt) 
\end{align}
which consists of two oscillatory terms, indicating two magnetically inequivalent stopping sites. 
Each term has a fractional contribution ($f_1$ and $f_2$) to the total oscillatory component $f_{\text{osc}}$ constrained by $f_1 + f_2 = 1$;
$f_1$ was found to be 0.59(1) at low temperature and ambient pressure and fixed at that value in all subsequent fits. 
The two oscillatory terms are  known as the Overhauser approximation~\cite{yaouanc_muon_2011} for incommensurate magnetic ordering with a field distribution experienced by the muon which is symmetric about some non-zero average field in the range $B_{i,\textrm{min}}\le B_i \le B_{i,\textrm{max}}$, with  
\begin{equation}
\label{eq:minmax}
B_{i,\textrm{avg}}=\frac{B_{i,\textrm{max}}+B_{i,\textrm{min}}}{2},\quad \Delta B_i=\frac{B_{i,\textrm{max}}-B_{i,\textrm{min}}}{2}
\end{equation}
and $J_0$ being the zeroth-order Bessel function of the first kind. 
Each term is damped at a respective rate $\Lambda_i$. 
A long-time exponential decay with the rate $\lambda_L$ constitutes the remaining fraction of the \ALRO\ response from the muons that experience a local field parallel to the initial muon spin orientation (on average 1/3 of the muons in an isotropic polycrystalline sample).
For ease of fitting, we only used a single $\lambda_L$ for both muon stopping sites. 
The fit parameters at low- and high-pressure regimes are listed in Table~\ref{tab:musr}.
We note that $f_{\text{osc}}$ is somewhat less than the expected value of 0.67 for an isotropic polycrystal, suggesting a small degree of preferred orientation in the pressed polycrystalline pellet.

\begin{table}[t]
 \centering
  \caption{\label{tab:musr}
  \fontfamily{phv}\selectfont\small{
   Fit parameters from Eq.~\ref{eq:Gmag} at ambient, low, and high pressures for $T\ll T_{\text{N}}$.
   Although the AFM transition appears sharper in the magnetization data, we use $\mu$SR fits (Fig.~S3b and S4b) to report \Tn\ values in this Table, so that all parameters are extracted from the same measurement.
  }}
  \begin{tabular}[width=0.46\textwidth]{|l|l|l|l|}
    \hline
    \rowcolor{lightgray}
    Pressure                    &    0~GPa    &  0.6~GPa   &  2.3~GPa   \\
    \hline  
    $T_{\text{N}}$ (K)          &   95.3(2)   &  95.4(8)   &   42.7(2)  \\ 
    $B_{1,\textrm{min}}$ (G)    &   1010(14)  &  1023(22)  &   14(3)    \\
    $B_{1,\textrm{max}}$ (G)    &   2134(14)  &  2155(22)  &   2010(3)  \\
    $B_{2,\textrm{min}}$ (G)    &   2193(14)  &  2203(13)  &   832(56)  \\
    $B_{2,\textrm{max}}$ (G)    &   2335(4)   &  2335(13)  &   1922(56) \\
    $\Lambda_1$ ($\mu$s$^{-1}$) &   2.6(4)    &    4(1)    &   0.2(1)   \\
    $\Lambda_2$ ($\mu$s$^{-1}$) &   2.2(2)    &    4.4(8)  &    7(3)    \\
    $f_{\text{osc}}$            &   0.58(1)   &   0.55(3)  &   0.48(3)  \\
    \hline
\end{tabular}
\end{table}

While \Tn\ is substantially reduced by the application of 2.3~GPa, consistent with the magnetization data, we find small changes in the local field parameters $B_{1,\textrm{max}}$ and $B_{2,\textrm{max}}$. 
Such modest changes of the upper limits on the local field (less than 20\%) could be accounted for by small changes of lattice parameters with pressure (Fig.~\ref{fig:XRD}b), which change the local field experienced by muons at the stopping sites (Fig.~\ref{fig:MUSR}e). 
The small change of local fields in $\mu$SR is consistent with the nearly unchanged magnetic moment under pressure in the Curie-Weiss analysis (Fig.~\ref{fig:MAG}e). 
These observations show the presence of robust local moments despite weakening of the magnetic order at high pressures, consistent with increasing $|K/J|$ ratio.

\section*{\label{sec:discussion}Discussion}
In previous studies,~\cite{banerjee_excitations_2018,tanaka_thermodynamic_2022} magnetic field has been used to melt the long-range order into a fluctuating regime in honeycomb lattices such as $\alpha$-\ch{RuCl3}.
Instead of changing the strength of $J$ or $K$ couplings, magnetic field enters the Hamiltonian as an external parameter (Zeeman term)~\cite{zhang_theory_2022,consoli_heisenberg-kitaev_2020}.
In contrast, pressure could tune the relative strength of competing interactions directly by changing orbital overlaps.
Despite theoretical proposals about using pressure as a powerful tuning parameter in Kitaev systems~\cite{yadav_strain_2018,bhattacharyya_maximized_2023}, an experimental verification has not been possible until now, because a small pressure is enough to induce a dimerization transition in both $4d$ systems (Ru$_2$ dimerization at 0.2~GPa in $\alpha$-\ch{RuCl3} and at 0.5~GPa in \ch{Ag3LiRu2O6})~\cite{bastien_pressure-induced_2018,stahl_pressure-tuning_2022,wang_pressure-induced_2018,takayama_competing_2022} and $5d$ systems (Ir$_2$ dimerization at 1.4~GPa in $\beta$-\ch{Li2IrO3})~\cite{veiga_pressure_2017}.
Remarkably, such a structural transition does not appear in \ALRO\ until 5~GPa, leaving a gap between the AFM (red) and dimerized (blue) phases in the phase diagram of Fig.~\ref{fig:MUSR}f. 

The emerging picture from our observations is a change of regime in \ALRO\ from a static AFM order to a dynamic spin liquid like state. 
Such a transition could be interpreted either within a $J$-$K$ model, as demonstrated in Fig.~\ref{fig:XRD}d, or within a $J_1$-$J_2$ model, although material-specific calculations are not available for this model. 
Moving forward, it will be helpful to get spectroscopic information from inelastic x-ray scattering and Raman scattering about the pressure induced dynamic regime near 4 GPa, and to search for evidence of quantum critical behavior by measuring temperature dependence of specific heat or NMR at low temperatures ($T<2$~K) near 4~GPa.
Such experiments, combined with material-specific calculations, could reveal the nature of the low-lying excitations in the gap between the red and blue phases in Fig.~\ref{fig:MUSR}f.

\pagebreak
\begin{methods}
\subsection{Material Synthesis.}
Polycrystalline samples of \ALRO\ were synthesized using a topotactic cation-exchange reaction from the parent compound \LRO\ following a previous publication~\cite{bahrami_first_2022}.
The structural and compositional quality of all samples were characterized at ambient condition with powder x-ray diffraction and energy dispersive x-ray spectroscopy.
The only impurity found was about 5\%\ pure silver inclusions.

\subsection{Magnetization measurements.} Magnetization of the powder sample was measured in a Quantum Design MPMS3 using a composite ceramic anvil cell~\cite{tateiwa_miniature_2011} with Daphne oil 7373 as the pressure-transmitting medium.
Pressure was determined from the superconducting transition of a lead manometer.
To achieve the maximum pressure of about 5.5~GPa, a pair of anvils with small culet sizes (1~mm) were used in runs 1, 2, and 3.
A small sample chamber with both diameter and thickness of 0.5~mm was drilled into the Be-Cu gasket.
To obtain data with higher quality for the CW fits, another pair of anvils with larger culets (1.8~mm) were used in run 4.
This time, the maximum pressure was about 2~GPa due to the larger sample chamber with both diameter and thickness of 0.9~mm.
In each run, magnetization of the empty cell was measured first as the background and subtracted from total signal. 
The small jumps near zero magnetization in Fig.~\ref{fig:MAG}f and Fig.~S1a,b are due to this subtraction. 

\subsection{Muon spin relaxation ($\mu$SR).}
The $\mu$SR experiments were performed at the Paul Scherrer Institute using the General Purpose Surface-Muon (GPS) and Decay-Channel (GPD) instruments on the "$\pi$M3"  and "$\mu$E1"  beamlines, respectively. 
Measurements on a pressed disk (12 mm diameter, 1 mm thickness) were made on GPS at ambient pressure using a gas flow cryostat between 110 and 6.5 K. 
Measurements in GPD at pressures of 0.57 and 2.29 GPa (as determined by an indium manometer) were made in He-flow cryostat using a piston-cylinder pressure cell~\cite{khasanov_perspective_2022} with Daphne oil 7373 as the pressure-transmitting medium. 
Data were analyzed using the MUSRFIT program~\cite{suter_musrfit_2012}.

\subsection{X-ray diffraction.}
X-ray diffraction (XRD) data were collected at the High Pressure Collaborative Access Team (HPCAT) beamline 16-BM-D of the Advanced Photon Source using diamond anvil cells (DAC) with a combination of full and partially perforated anvils to reduce x-ray attenuation. 
Anvil culet diameter was 300 $\mu$m. 
Rhenium gaskets were pre-indented to a thickness of 50~$\mu$m, and a 180~$\mu$m-diameter sample chamber was laser drilled at the center of the indentation. 
Fine powder (~5 $\mu$m) of \ALRO\ together with ruby and gold manometers were loaded into the sample chamber filled with Ne pressure medium. 
The entire sample chamber was rastered over the $25\times25~\mu\text{m}^2$ area of the 30 keV X-ray beam to improve powder averaging on the CCD detector.  
Measurements were carried out at both ambient and low temperature (83 K). 
2D XRD images were integrated over $2\pi$ using Dioptas software~\cite{prescher_dioptas_2015} and the integrated diffractograms were Le Bail fitted using Jana2020~\cite{petricek_jana2020_2023}. 
Pressure-dependent lattice parameters were extracted and 2nd order Vinet and Birch-Murnagham equations of state were both fitted using EoSFit~\cite{gonzalez-platas_eosfit7-gui_2016}.

\subsection{DFT calculations.}
Structural optimization and electronic structure calculations at high pressures were performed using the QUANTUM ESPRESSO and Wannier90 codes~\cite{giannozzi2009quantum,giannozzi2017advanced,Pizzi2020} with the experimental crystallographic information as the input.
To evaluate the wavefunctions in the supplementary information (Table~S1), we first used Quantum ESPRESSO and Wannier90 codes to compute the electronic structure using experimental lattice parameters from our XRD measurements under pressure.
Then, a tight-binding model was constructed for an individual RhO$_6$ cluster, defined by real-space hopping parameters extracted from DFT. 
The orbital information were calculated from a Hartree-Fock mean-field model. 

\subsection{Neutron diffraction.}
Neutron powder diffraction (NPD) was performed on 2 grams of polycrystalline \ALRO\ using the HB-2A powder diffractometer and the HB-1A~\cite{aczel_revisiting_2019} Triple-Axis-Spectrometer (VERITAS) at the High Flux Isotope Reactor (HFIR) at Oak Ridge National Laboratory (ORNL).  
On HB-2A, the sample was loaded into a 5~mm diameter Al can to give an overall neutron transmission of 77.67\%. 
We used collimations of open-21'-12' with a wavelength of 2.41$\textrm{\AA}$. 
On HB-1A the sample was loaded into an annular can with 1~mm annulus and resulting neutron transmission of 90.38\%. 
We used collimations of 40’-40’-40’-80’ with a fixed incident energy of 14.5~meV. 
FULLPROF~\cite{rodriguez-carvajal_recent_1993} was used for Rietveld refinements of crystal structures and computing predicted magnetic diffraction patterns to compare with experimental data.

\end{methods}



\pagebreak
\section*{References}
\bibliographystyle{naturemag} 
\bibliography{Sakrikar_19june2024}


\pagebreak
\begin{addendum}
 \item[Acknowledgments]
 The authors thank L.~Hozoi for fruitful discussions.
 The work at Boston College was supported by the U.S. Department of Energy, Office of Basic Energy Sciences, Division of Physical Behavior of Materials under Award No. DE-SC0023124.
 The work in Augsburg was funded by the Deutsche Forschungsgemeinschaft (DFG, German Research Foundation) - TRR 360 - 492547816. 
 Bin Shen acknowledges the financial support of Alexander von Humboldt Foundation.
 K.W.P. and Q.W. were supported by the U.S. Department of Energy, Office of Basic Energy Sciences, under Grant No. DE-SC0021223.
 This work is based in part on experiments performed at the Swiss Muon Source S$\mu$S, Paul Scherrer Institute, Villigen, Switzerland. 
 Neutron scattering experiments were carried out at the High Flux Isotope Reactor and Spallation Neutron Source, a DOE Office of Science User Facility operated by the Oak Ridge National Laboratory.  
 Y.R. and X.H. acknowledge support from the National Science Foundation under Grant No. DMR-1712128.
 E.P. and R.J.H acknowledge support from DOE-SC (DE-SC0020340), DOE-NNSA (DE-NA0003975), and NSF (DMR-2118020 and DMR-2119308). 
 Operations of HPCAT (Sector 16, APS, ANL) are supported by DOE-NNSA’s Office of Experimental Sciences.
 Work at the Advanced Photon Source was supported by the U.S. Department of Energy Office of Science, Office of Basic Energy Sciences, under Award No. DE-AC02-06CH11357.

 \item[Author Contributions] P.S., C.W., E.M.K., R.G., R.K., H.L., and M.J.G performed $\mu$SR experiments. 
 B.S., K.W.F., P.G., and A.T. performed magnetization measurements. 
 E.D.T.P., G.F., R.J.H., and D.H. performed x-ray diffraction. 
 Q.W., S.A.C., A.A.A., and K.W.P. performed neutron diffraction. 
 F.B. synthesized the material. 
 X.H. and Y.R. performed theoretical calculations. 
 F.T. conceptualized and coordinated the project. All authors participated in the writing process.  
 \item[Inclusion and Ethics] Contributions from all authors including local scientists at the national and international labs are properly acknowledged in this work.
 \item[Competing interests] The authors declare no competing interests.
 \item[Data availability] Once the paper is accepted for publication, we will provide DOI for the data repository.
 \item[Supplementary information] is available online including the crystallographic information file (CIF).
 \item[Correspondence and requests for materials] should be addressed to F.T. via email fazel.tafti@bc.edu
\end{addendum}

\end{document}